\documentclass[%
reprint, % preprint
 amsmath,amssymb,
 aps, physrev,
]{revtex4-2}

\usepackage{graphicx}% Include figure files
\usepackage{dcolumn}% Align table columns on decimal point
\usepackage{bm}% bold math
\usepackage{siunitx}

\usepackage{xcolor}

\begin{document}
\preprint{APS/123-QED}

% \title{\textbf{Measurement of deformation state inside the thin material ligament connecting two crack steps} 
% }% 

\title{\textbf{Falling through the cracks: energy storage along segmented brittle crack fronts
} 
}% 
% 3D structure as elastic energy reservoir in brittle fracture
%% ligament structure as an energy reservoir 
%% Extra energy available to drive fracture growth
%% and again how local affect global

\author{Xinyue Wei} 
\author{John M. Kolinski}%
\email{Contact author: john.kolinski@epfl.ch}
\affiliation{%
 Institute of Mechanical Engineering, School of Engineering, Ecole Polytechnique Fédérale de Lausanne (EPFL), 1015 Lausanne, Switzerland
}%

\date{July 2025}% It is always \today, today,

\begin{abstract}
During brittle crack propagation, a smooth crack front curve frequently becomes disjoint, generating a stepped crack and a material ligament that unites the newly formed crack fronts. These universal features fundamentally alter the singular field structure and stability of propagating cracks; however, a quantitative analysis of their mechanics is lacking. Here, we perform in-situ 3D measurements to resolve the deformation field around stepped cracks, and crucially, within the ligament feature. The 3D kinematic data are obtained by scanning a thin laser sheet through the brittle hydrogel samples, while recording the scattered intensity from the embedded tracer particles. We find that the ligament concentrates the strain energy density, and moreover, the apparent fracture energy increases proportionally to the strain energy within the ligament.
\end{abstract}

\maketitle
\begin{figure}[t]
    \centering
\includegraphics[width=1\linewidth]{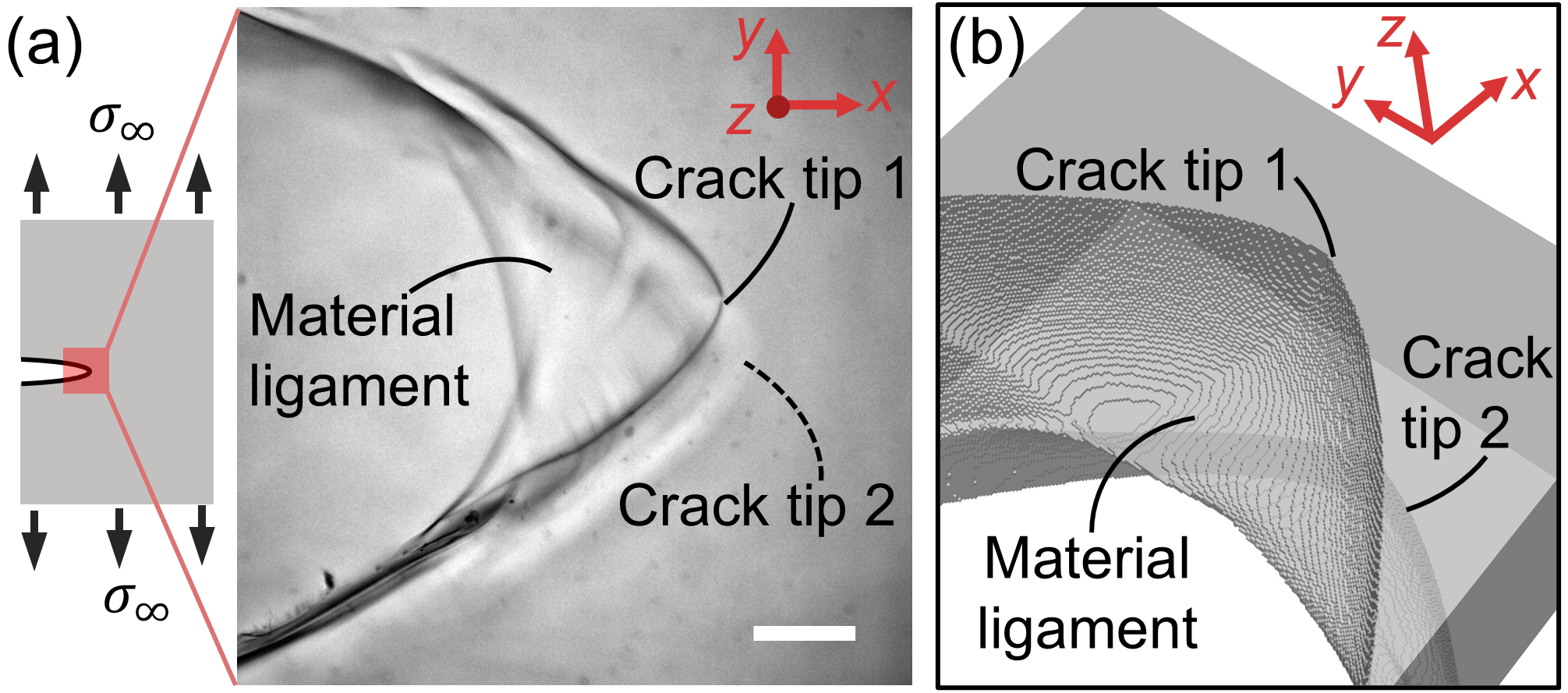}
    \caption{Images of a stepped crack and material ligament. (a) Left: Schematic of a mode I crack with the crack tip region highlighted. Right: Bright field micrograph of a stepped crack in brittle hydrogel during its propagation. The crack has two crack tips -- crack tip 2 is at the tip of the blurry feature out of the focal plane. There is a ligament of material that connects the two crack steps and falls behind the crack tips. The scale bar is \SI{100}{\micro m}. (b) 3D graph of (a).}
    \label{fig:stepped crack and ligament}
\end{figure}

Cracks routinely break their planar symmetry, generating a universal three-dimensional structure consisting of disjoint crack front segments joined by material ligaments~\cite{sommer_formation_1969, lazarus_crack_2001, tanaka_discontinuous_1998,baumberger_magic_2008,pons_helical_2010, ronsin_crack_2014,   pham_formation_2017, kolvin_topological_2018, wang_how_2022,steinhardt_how_2022, wang_dynamics_2023, wang_size_2024}. The shape of a typical propagating crack with broken planar symmetry is illustrated in Fig.~\ref{fig:stepped crack and ligament}, where the crack front is spatially fragmented along the sample thickness $z$. Between the front segments, the ligament lags behind the advancing cracks. The observed planar symmetry breaking can be caused by material heterogeneity~\cite{steinhardt_how_2022, wang_how_2022, wei_complexity_2024} or far-field loading conditions~\cite{lazarus_crack_2001, leblond_theoretical_2011, chen_crack_2015,vasudevan_configurational_2020, lubomirsky_topological_2024, lebihain_crack_2023}; these non-planar cracks are inherently incompatible with the linear elastic fracture mechanics theory (LEFM)~\cite{freund_lambert_ben_dynamic_1998, anderson_ted_l_fracture_2017}, and thus we lack a predictive theory for their physics. While crack front perturbation theories have attempted to identify the onset of this instability \cite{xu_analysis_1994, leblond_theoretical_2011, pham_growth_2016} and perturbations to the fracture energy ~\cite{gao_somewhat_1987, perrin_disordering_1994,movchan_perturbations_1998, ramanathan_onset_1998, adda-bedia_dynamic_2013,leblond_out--plane_2016, fekak_crack_2020, lebihain_size_2023, kolvin_dual_2024, kolvin_comprehensive_2024} model dynamics after its onset, the 3D mechanical state of a ligament remains unknown, making it impossible to predict the trajectory of a non-planar crack.

Here, we probe the 3D kinematics around a stepped crack in a brittle hydrogel material with in-situ imaging via light sheet microscopy to capture the deformation field within the material ligament. Using the 3D kinematic data, we evaluate the mechanical state inside and around the ligament feature, and find that both the elastic strain energy and the material stretch is enhanced and localized within the ligament feature. Upon evaluating the integrated strain energy stored in the ligament, and comparing it with the far field strain energy release rate, we find that these quantities are proportional to one another, and thus identify the key governing role of the local complex crack front geometry in the global failure outcome.

As a proxy brittle material, we use polyacrylamide hydrogels with the 13.8\% w/v acrylamide monomer and 2.67\% of which is the bis-acrylamide crosslinker. The hydrogel is transparent, brittle, and the constitutive behavior can be described as incompressible Neo-Hookean solid with the shear modulus of \SI{35}{kPa}. The same hydrogel is widely used in previous studies on brittle failure~\cite{livne_breakdown_2008, livne_near-tip_2010, goldman_boue_failing_2015, kolvin_nonlinear_2017, kolvin_topological_2018, li_crack_2023,wei_complexity_2024}, and its fracture mechanics have been carefully characterized~\cite{yang_polyacrylamide_2019, liu_polyacrylamide_2019, wang_polyacrylamide_2021, wang_polyacrylamide_2023, kim_polyacrylamide_2022, wang_polyacrylamide_2023}. To image the gel, a small portion of the acrylamide monomer is fluorescently labeled; to monitor material displacement, \SI{1.1}{\micro m} diameter polystyrene colloids are embedded within the gel~\cite{benkley_estimation_2023}. The samples are prepared with \SI{1}{cm}$\times$\SI{3}{cm}$\times$\SI{200}{\micro m} shape, and placed in a 90.3\% glycerol-water solution until the solvent exchange reaches equilibrium ensuring a close match of the gel's optical index with the surrounding liquid during the experiment. Details of the material preparation can be found in the Supplemental Materials. 

To perform the fracture test, an edge crack is cut perpendicular to the long edge of the sample at its center, and the sample is placed in grips along the short edges such that the crack is centered in our optical system's field of view. Remote tensile stretches are applied from the grips in small increments using a controlled motorized stage.

\begin{figure}[t]
    \centering
    \includegraphics[width=\linewidth]{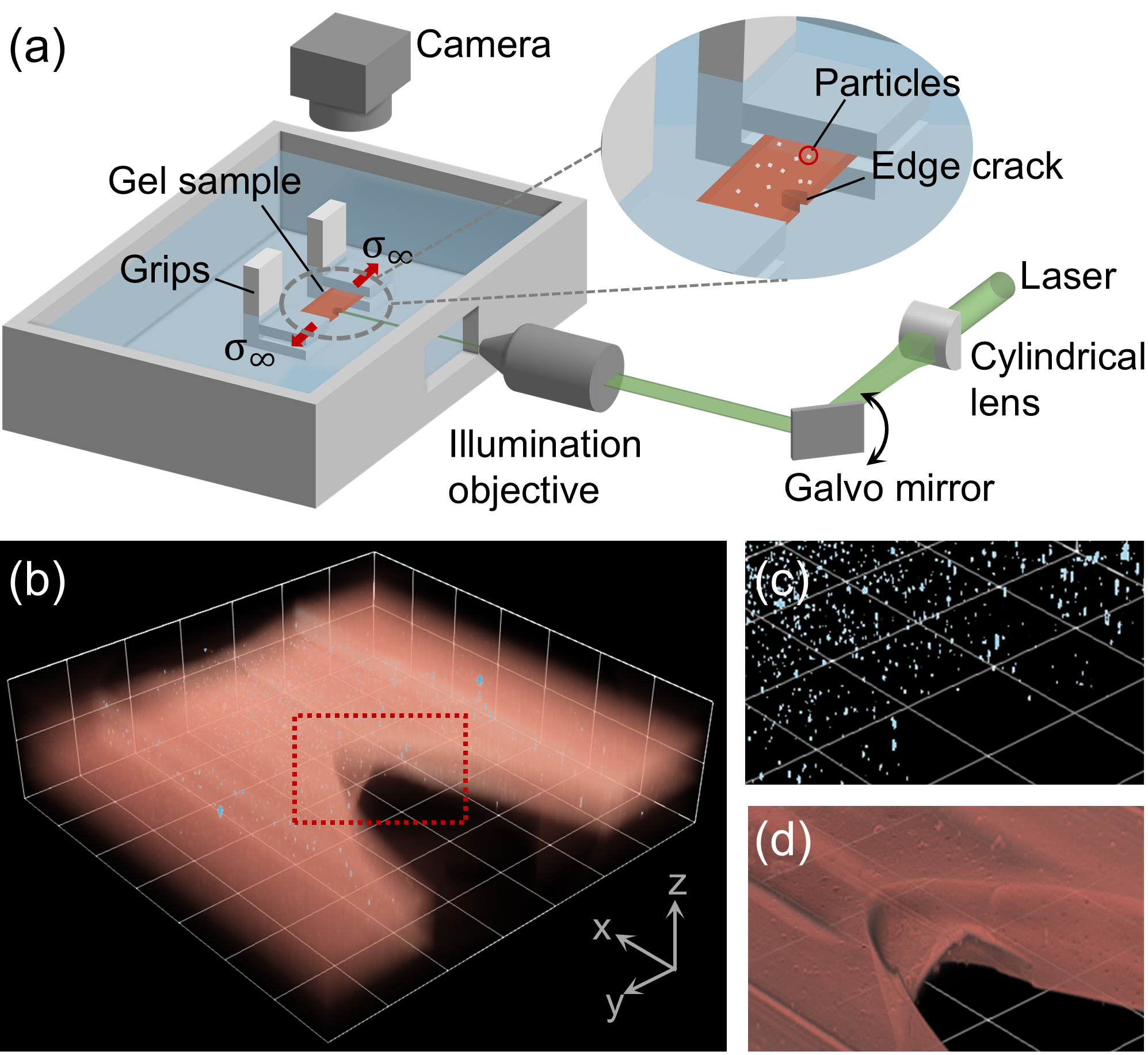}
    \caption{Image collection and volumetric rendering. (a) Schematic of the experimental setup. A laser sheet $\sim$\SI{8}{\micro m} in thickness is generated by a cylindrical lens and scanned through the sample thickness by a galvo mirror. The fluorescently labeled hydrogel samples are embedded with tracer particles. The sample (\SI{1}{cm}$\times$\SI{3}{cm}$\times$\SI{200}{\micro m}) has a pre-cut edge crack and remote tensile loading is applied at the sample grips with small stretch increments controlled by a motorized stage. 3D volumetric images are captured by the camera throughout the loading cycle. (b) A typical example of a 3D image stack for a stepped crack. The dashed box circumscribes the magnified views in (c) and (d). The grid spacing is \SI{200}{\micro m}. (c) Visualization of segmented particle locations near the crack tip. (d) Visualization of gel geometry.}
    \label{fig:setup}
\end{figure}

To image the in-situ 3D deformation around the material ligament, we adapted a laser sheet imaging platform to capture volumetric data as the sample is loaded. The experimental setup is depicted schematically in Fig.~\ref{fig:setup}(a). The optical path for the light sheet is adapted from OpenSPIM~\cite{pitrone_openspim_2013}, where a driven galvanometric mirror is used to displace the light sheet along the optical axis of the imaging path. To generate the light sheet, a laser beam with a wavelength of \SI{625}{nm} is first shaped into a wide planar beam using a cylindrical lens and then focused to a thin laser sheet into the gel sample using the illumination objective. The thickness of the beam is $\sim$\SI{8}{\micro\meter} and remains approximately constants across our field of view (\SI{1.4}{mm}$\times$\SI{1.4}{mm}), enabling both good $z$ sectioning, and a sufficient lateral extent in the $xy$ plane.

For each stretch increment, the fluorescent and scattered light are imaged onto the sensor of the camera (Hamamatsu C13440, 2048$\times$2048 pixels, 16 bit). To enable the volumetric reconstruction, a series of 2D images is recorded as the galvo steers the light sheet through the sample thickness in steps of \SI{2}{\micro m}; the total sampled thickness is \SI{400}{\micro m}, corresponding to a stack of 200 images. The depth of field of the imaging system ensures that each image is in focus. A sample 3D rendering, where the images are processed to segment the image content, is shown in Fig.~\ref{fig:setup}(b). Each image stack consists of information in two intensity bands: one band contains light scattered from the embedded tracer particles, and the second band corresponds to the fluorescent emission from the hydrogel, as shown in Fig.~\ref{fig:setup}(c)\&(d), respectively. The two intensity channels provide high-contrast signals for particle tracking on the one-hand, and a sharp contrast between the gel and the environment on the other hand. By acquiring such volumetric data throughout the loading cycle -- from an initially undeformed configuration to the fully opened state -- we capture both the topology of the in-situ crack front and crack faces, as well as the displacement inside the material around the crack tip. The segmentation and particle tracking workflow is explained in the Supplemental Material. 

\begin{figure}[t]
    \centering
    \includegraphics[width=\linewidth]{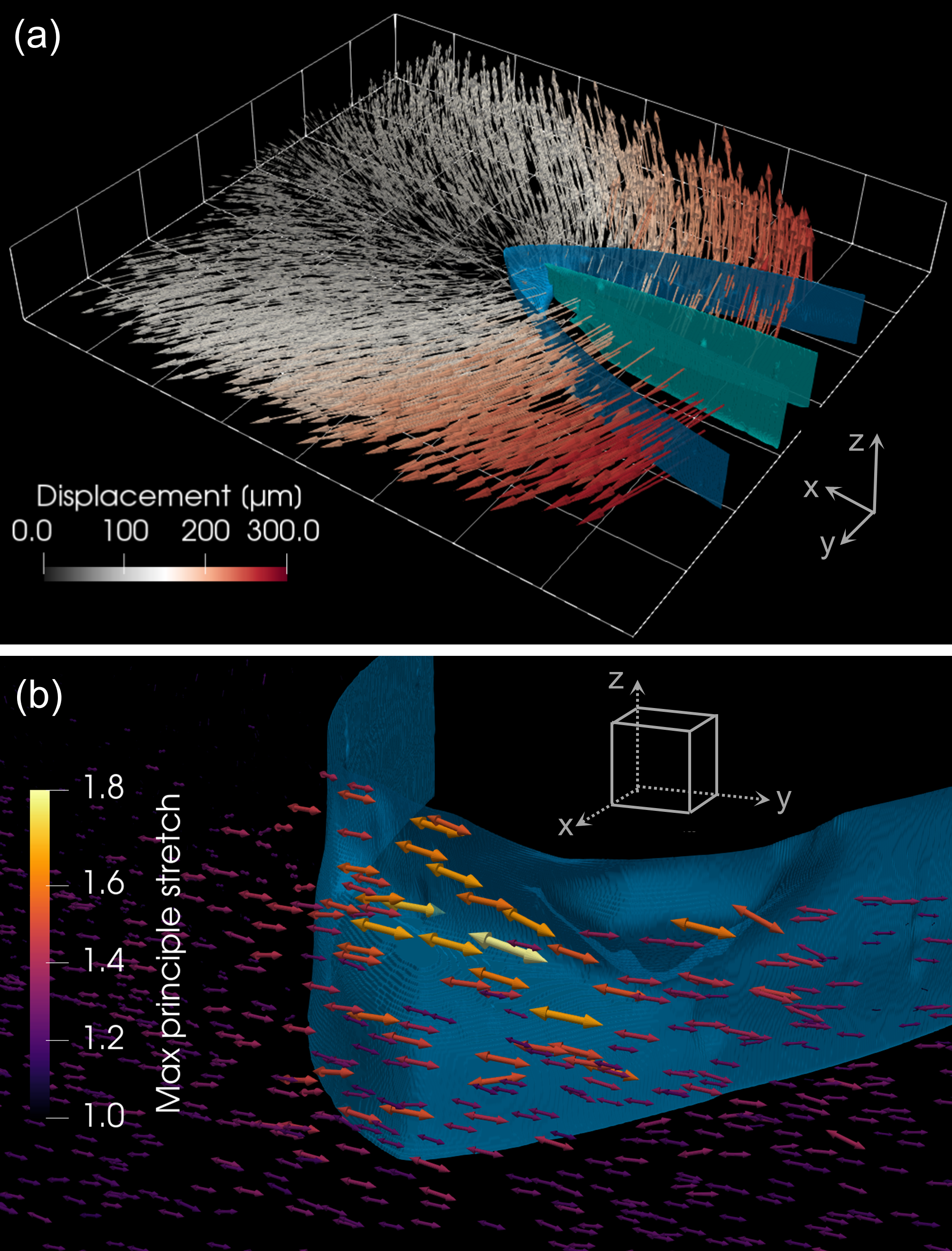}
    \caption{Kinematic field structure near the crack tip. (a) The particle displacements are graphed with the fracture surface in the rest state (cyan) and the deformed state (blue). The grid spacing is \SI{200}{\micro m}. (b) The maximum principle stretch is highest inside the material ligament and aligns with the ligament direction. The cube shown on the coordinate axes has sides of length \SI{50}{\micro m}.}
    \label{fig:stretch}
\end{figure}

The displacement field from both the CTOD measurement and from the particle tracking follow a typical pattern due to the symmetric remote tensile loading. As the crack opens from the rest to the deformed state, both the far-field CTOD and the magnitude of the displacement vectors reflect the parabolic trend predicted by LEFM as shown in Fig~\ref{fig:stretch}(a). The far-field symmetry is significantly altered at the crack tip due to the rich 3D structure of the step. The displacement field shows strong variation throughout the bulk, suggesting kinematic gradients that emerge in responds to the applied loading.

To fully characterize the kinematics of the bulk around the crack tip, we estimate the deformation gradient tensor $\mathbf{F}$ with the current and reference position vectors of the particles as described elsewhere~\cite{benkley_estimation_2023}. The deformation gradient is defined as $\mathbf{F} = \frac{d\mathbf{x}}{d\mathbf{X}}$, where $\mathbf{X}$ and $\mathbf{x}$ denote the initial and deformed positions, respectively. The deformation gradient is evaluated on every particle location without interpolating on a grid, ensuring fidelity of the data in features near the free surface, including the ligament. To separate the rigid body motion from the deformation, we apply polar decomposition to $\mathbf{F}$, and thus obtain the rotation tensor $\mathbf{R}$ and the stretch tensor $\mathbf{U}$ or $\mathbf{V}$ through $\mathbf{F} = \mathbf{R}\mathbf{U} = \mathbf{V}\mathbf{R}$. These data provide insight into the state of deformation throughout the bulk of the material for fully loaded cracks.

% \begin{figure}
%     \centering
%     \includegraphics[width=0.6\linewidth]{FiG_{\mathrm{c}}rack_front_projection.png}
%     \caption{Crack front curve in a stepped crack and its projections on xz and yz plane}
%     \label{fig:enter-label}
% \end{figure}

What happens inside the ligament? To address this question, we scrutinize the local stretch tensor obtained from the 3D displacement field, and calculate the eigenvalues and eigenvectors of the stretch tensor $\mathbf{V}$, i.e. the principle stretches. The greatest principle stretch vector characterizes the orientation of the greatest deformation within the body; the regions of greatest principal stretch are located within the ligament, and the direction of these maximum principal stretches is aligned with its orientation, as shown in Fig.~\ref{fig:stretch}(b). Surprisingly, the peak principal stretch does not occur near the upper or lower crack tips, but instead localizes within the ligament itself. This concentration of maximum principal stretch suggests that the ligament acts as a mechanical bottleneck for the fracture process: it must undergo significant extension and ultimately fail before the adjacent crack segments can propagate further. This implies that the ligament plays a critical role in regulating the propagation of the crack front, functioning not merely as a passive connector, but as an important structural feature that resists fracture. In this sense, the ligament effectively ``holds" or ``shares" the applied far-field stretch, temporarily stabilizing the crack front while storing a significant portion of the strain energy. In addition, the stress field also shows a concentration within the ligament, however, this localized stress does not significantly alter the overall stress singularity behavior on and off the ligament plane. Details for the calculation of the stress field from the kinematic data and further analysis are discussed in the Supplemental Material.

\begin{figure}[t]
    \centering
    \includegraphics[width=\linewidth]{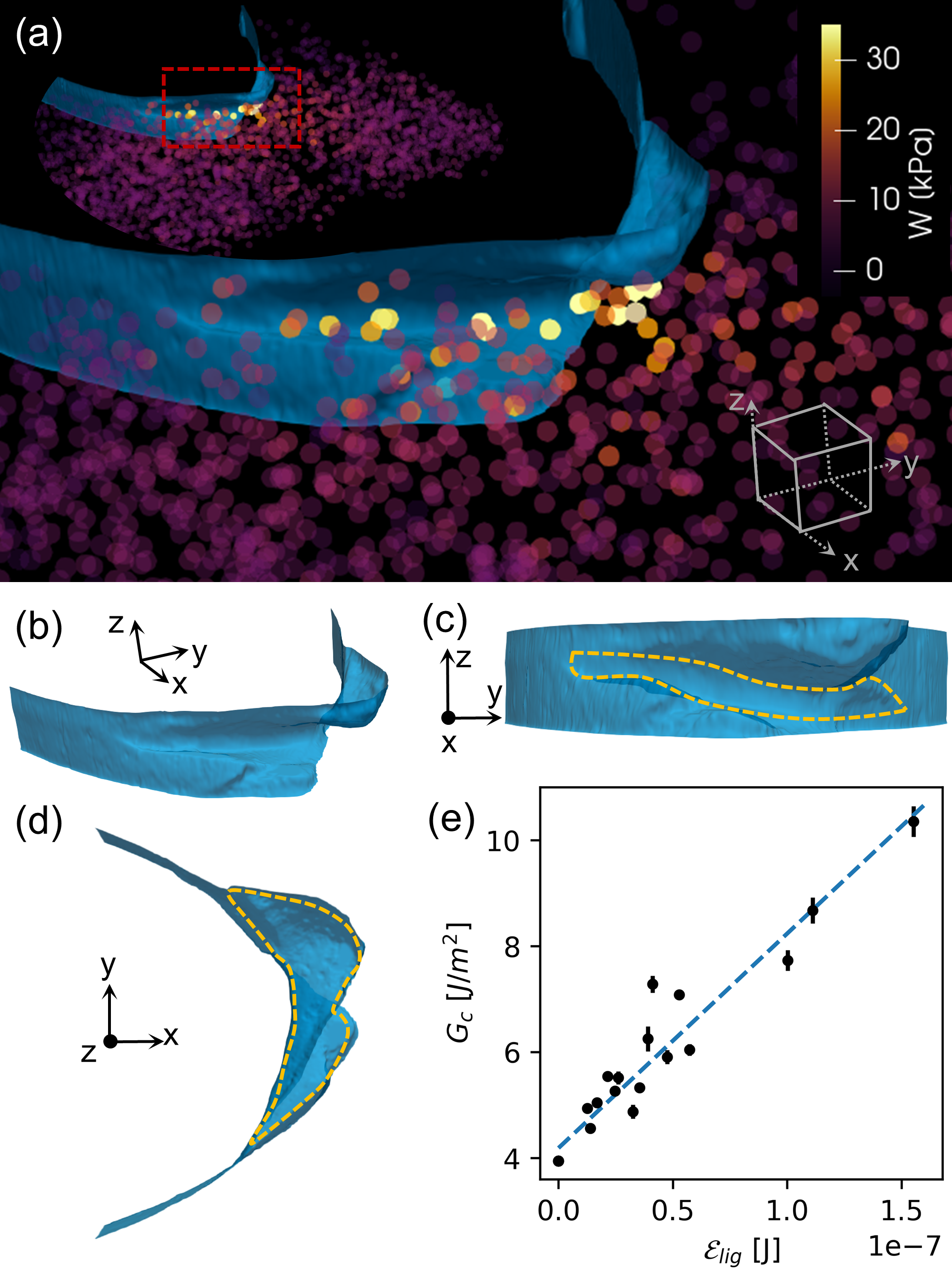}
    \caption{Strain energy between the cracks. (a) The strain energy density $W$ concentrates inside the ligament. The sides of the box are of length \SI{50}{\micro m}.(b-d) The crack surface in (a) is shown with its projections on the $yz$ plane (c) and the $xy$ plane (d). The ligament region is marked by dashed yellow curves. (e) The critical strain energy release rate $G_{\mathrm{c}}$ increases linearly with the ligament strain energy $\mathcal{E}_{\mathrm{lig}} = \int_{V_{\mathrm{lig}}} W dV$.}
    \label{fig:W}
\end{figure}

The concentration of the strain and stretch inevitably lead to the concentration of the strain energy density $W$. With the incompressible Neo-Hookean constitutive model, we can calculate the strain energy density at each of the tracked particles $W = \frac{\mu}{2} (I_1 - 3)$, where $\mu$ is the shear modulus and $I_1$ is the first principal invariant (trace) of the left Cauchy-Green deformation tensor $\mathbf{B = FF^T = V^2}$. In analyzing the spatial dependence of $W$, we identify a clear and localized peak within the ligament region, as shown in Fig.~\ref{fig:W}(a), indicating that the elastic energy density is significantly higher than in the surrounding matrix. Physically, the ligament material is locally storing a large amount of elastic energy per unit volume. 

How much strain energy is stored inside the ligament? With our definition of ligament particles (see Supplemental Material), we can now calculate the total energy stored in the ligament region by integrating the strain energy density over its volume. We approximate the integrated energy by $\mathcal{E}_{\mathrm{lig}} = \int_{V_{\mathrm{lig}}} W dV \approx \sum_{i} W_i (\frac{4\pi}{3} r_i^3)$, where $W_i$ and $r_i$ are the strain energy density and the radial weighting of the $i$th particle, respectively; the summation is carried out over all particles identified within the ligament region. 

To evaluate the impact of the elastic energy stored in the ligament on the overall fracture resistance of the crack, we compare it with the critical strain energy release rate $G_{\mathrm{c}}$, as measured from parabolic fitting of the far filed CTOD data~\cite{wei_complexity_2024}. We find that $G_{\mathrm{c}}$ increases linearly with the energy stored inside the material ligament, as shown in Fig.~\ref{fig:W}(c), and thus
\begin{equation}
    G_{\mathrm{c}} \propto \mathcal{E}_{\mathrm{lig}}.
\end{equation}

% \begin{equation}
%     G_{\mathrm{c}} = \alpha \int_{V_{\mathrm{lig}}} W dV + b, 
% \end{equation}
% where $\alpha$ is the slope and $\Gamma_{clean}$ is the intersect. 

Why is $G_{\mathrm{c}}$ proportional to $\mathcal{E}_{\mathrm{lig}}$? The ligament strain energy $\mathcal{E}_{\mathrm{lig}}$ quantifies the total strain energy stored within the volume of the material ligament, while the critical strain energy release rate $G_{\mathrm{c}}$ quantifies the amount of energy that must be released per unit crack advance per unit sample thickness. In other words, $G_{\mathrm{c}}$ represents the resistance of the crack, or how much energy is required to drive the crack as a whole. The relationship identified between $G_{\mathrm{c}}$ and $\mathcal{E}_{\mathrm{lig}}$ reveals that there is ``extra" energy stored in the material ligament when compared to a smooth crack front. As the ligament accumulates more strain energy, the crack must overcome a larger energetic barrier in order to propagate. The more complex the ligament geometry, the more energy can be stored and subsequently released during fracture, consistent with prior observations of $G_{\mathrm{c}}$ and crack front complexity~\cite{wei_complexity_2024}. Thus, the material ligament plays a crucial role in the 3D energy storage and dissipation, and acts like an energy reservoir. 

% How do we understand the slope and the intersect in the linear proportionality? 
The linear fit of $G_{\mathrm{c}}$ versus $\mathcal{E}_{\mathrm{lig}}$ yields a slope of $3.84\times10^7 \unit{m^{-2}}$ and an intercept of \SI{4.36}{J/m^2}. The intercept corresponds to the baseline fracture energy for a clean crack, where the ligament volume is zero. The ligament energy $\mathcal{E}_{\mathrm{lig}}$ represents how much energy is stored inside the material ligament. However, it is unknown how this stored energy will be released. The slope $3.84\times10^7 \unit{m^{-2}}$ reflects the unit area of over which $\mathcal{E}_{\mathrm{lig}}$ will be released.

Here, we have analyzed the mechanical fields near the tip of a complex crack in a brittle hydrogel material using light sheet microscopy to collect imaging data of scattered and fluorescent light. We find that the universal ligament feature localizes high values of both the stretch and the strain energy density. The far field strain energy release rate is proportional to the quantity of elastic energy stored in the ligament feature. 

The linear relationship between $G_{\mathrm{c}}$ and $\mathcal{E}_{\mathrm{lig}}$ rigorously supports the prior energetic analysis, where the 3D structure of the complex features of the crack front restore energy balance~\cite{wang_how_2022, wei_complexity_2024}. In those studies, energy balance was recovered by incorporating the increased crack front length and the geometry of the material ligament. Our results extend and support this picture by the direct measurement of the energy stored in the ligament, and by showing that this energy correlates linearly with the additional fracture energy required for propagation. This not only validates the generalized 3D energy balance argument proposed in \cite{wang_how_2022}, it establishes a concrete role for material ligaments in the energy balance that governs fracture resistance.

Our results suggest that the local dynamics of such geometric features can be quantitatively analyzed, and may be used in the analysis of cracks where the complex 3D field structure is represented as a local increase in the fracture toughness~\cite{leblond_theoretical_2011, lebihain_size_2023} or an effective cohesive zone~\cite{lazarus_mode_2020}. However, these universal features are known to have systematic -- rather than random -- dynamics~\cite{tanaka_discontinuous_1998, baumberger_magic_2008, pons_helical_2010,ronsin_crack_2014, kolvin_topological_2018, wang_dynamics_2023, steinhardt_geometric_2023, wang_size_2024,cochard_propagation_2024, wei_complexity_2024}, that are generated from the local structure, and thus the fully 3D field structure must ultimately be accounted for to improve the predictive modeling of crack growth in 3D.

 The energy dissipation within the ligament is rather complex. The slope of the linear proportionality between $G_\mathrm{c}$  and $\mathcal{E}_\mathrm{lig}$ characterizes how the energy stored in the ligament is released as the crack advances. A naive attempt to assign and area over which the energy stored in the ligament is released fails to obtain quantitative agreement with $G_\mathrm{c}$ as described in the Supplemental Material; this suggest that further analysis is required to understand why $G_\mathrm{c}$ is proportional to $\mathcal{E}_\mathrm{lig}$. To fully understand this complex local fracture process, direct observation of the ligament breakage in 3D, during the propagation of a stepped crack, is essential. Future work on the dynamic evolution of the ligament during propagation will be able to provide us with insight into why the energy dissipated at the ligament is universal as the measured slope suggests.
 
 % It is important to recognize that not all of the stored energy is necessarily released simultaneously or uniformly, as discussed in the Supplemental Material. The ligament may rupture progressively, with some regions failing earlier than others, leading to a temporally and spatially heterogeneous energy release process. In this context, the slope is a key parameter that can be interpreted as an average efficiency of energy release during fracture. 

% For a baseline fracture energy of \SI{5}{J/m^2} and a sample thickness of \SI{200}{\micro m}, the strain energy dissipation to propagate the crack for \SI{100}{\micro m} is \SI{1e-7}{J}, which means that when $\mathcal{E}_{\mathrm{lig}}=$\SI{1e-7}{J}, the apparent fracture energy of the crack will double. 

% \begin{itemize}
%     \item propagating singular fields
%     \item symmetry breaking and topology
%     \item experimental method allows us to probe singular field directly
%     \item universality of K in the far field
%     \item microscale to macroscale
%     \item methodological relevance for exploring complex systems including soft materials, contact lines and glassy systems
% \end{itemize}

Our findings in this work provide quantitative insight into how topological features, such as material ligaments, influence the energy distribution and fracture resistance in soft brittle materials. By fully resolving the 3D deformation field around stepped cracks, this work contributes to the fundamental understanding of how 3D geometry affects the physics of global fracture. This localized energy storage challenges the conventional picture of planar cracks, and highlights the need for a three-dimensional energetic framework when analyzing crack dynamics.

With our measurements, we probe the mechanical state of the material near a crack tip when that crack has broken planar symmetry, and generated a fundamentally different crack front topology, that is seemingly universal in the failure of brittle solids~\cite{tanaka_discontinuous_1998, kolvin_topological_2018,steinhardt_geometric_2023, wei_complexity_2024}. Our fully resolved 3D measurements bridge the local structure, where dissipation occurs at the complex crack front, to the global forcing in the far field; this method can be readily adapted in other settings such as soft, active and biological materials~\cite{doi_masao_soft_2013, angelini_glass-like_2011}, contact lines~\cite{de_gennes_capillarity_2003}, and glassy systems~\cite{weeks_three-dimensional_2000}, where the dynamics spans across spatial scales. Whereas we find the broken symmetry significantly alters the local field structure at the tip of the crack, we recover the universal $K$ field away from the crack tip, albeit with a modified value of $K$, as described in the Supplemental Material. This approach enables the direct experimental characterization of propagating singularities, where the local dissipation governs the global outcome~\cite{jens_g_eggers_singularities_2023}.

\textit{Acknowledgment} -- The authors gratefully acknowledge Chenzhuo Li for extensive discussions. The authors acknowledge the enlightening discussions with participants of the CECAM workshop on ``3D cracks and crack stability". The authors acknowledge the support of the Swiss National Science Foundation grant No.197162.

\bibliography{references.bib}
% \bibliography{references-2.bib}
\end{document}